# Impedance spectroscopy and conduction mechanism of a $BiFe_{0.95}Mn_{0.05}O_3$ thin film

S. Yousfi, M. El Marssi, H. Bouyanfif*

Laboratoire de Physique de la Matière Condensée, Université de Picardie Jules Verne,

33 Rue Saint Leu, 80039 Amiens, France

## Abstract

Dielectric response and conduction mechanism were investigated for a multiferroic $BiFe_{0.95}Mn_{0.05}O_3$ epitaxial thin film. A contribution from a thermally activated interface (0.37 eV) and the bulk of the film on the dielectric response were observed through the comparison between experimental results and equivalent circuit model. The low frequency interface relaxation signatures strongly suggest a Maxwell-Wagner space charge origin. The alternative current conductivity deduced from the model follows a power law frequency dependence suggesting a polaronic hopping mechanism while the low frequency limit is in perfect agreement with the direct current conduction mechanism. The current-voltage characteristics were indeed correlated with Schottky-Simmons interface limited transport with activation energy of 0.36 eV, close to the one extracted from the impedance analysis. Such analysis of the electrostatic landscape and dielectric behaviour may help to further understanding the anomalous photo-induced properties in the $BiFeO_3$ system.

*corresponding author: houssny.bouyanfif@u-picardie.fr

## Introduction

$BiFeO_3$ (BFO) is the most studied multiferroic system due to the room temperature coexistence of a ferroelectric and antiferromagnetic order. The room temperature multiferroic nature of the BFO allows indeed broad range of applications in various domains such as for example electronic, optoelectronics, energy storage and information storage [1-4]. BFO and BFO based solid solutions scientific and technological importance are emphasized by the many number of articles published (see [1- 12] and references therein). BFO bulk adopts rhombohedral structure (R3c space group) and a pseudocubic lattice parameter of about 3.96 Å. The Curie and Neel temperature of BFO are respectively 1100 K and 640 K and the spontaneous polarization lies along the [111] pseudo-cubic direction (Ps = 100 μC/cm$^2$). Despite a large number of investigations on BFO in bulk or nanostructures, dielectric studies are scarce in epitaxial and single crystal like BFO thin film. A large variety of dielectric investigations were indeed performed on BFO doped thin films made by sol gel or chemical method of synthesis and with textured growth. Nevertheless, these works have importantly shown the possibility to improve the dielectric response via chemical substitution [5,6]. Dielectric losses can be therefore decreased by a suitable chemical substitution and this is of importance for magnetodielectric coupling optimization. Amongst the very few investigations in epitaxial singly crystal-like thin film, Peng et al. have revealed coexisting bulk and interface contributions in BFO epitaxial thin film grown on Nb doped (001) oriented $SrTiO_3$ substrate [7]. Considering the very strong sensitivity of dielectric properties to strain and type of electrodes, it would be important to screen the dielectric properties and we present therefore the impedance spectroscopy analysis of a different heterostructure (ITO/BFO/SRO/STO) to those observed in the literature. We describe in this article the impedance spectroscopy analysis of a highly insulating epitaxial BFO thin film grown on $SrRuO_3$ buffered (001) oriented $SrTiO_3$ substrate in parallel plate capacitance geometry.

To associate different immittance between them, the following equations (1), (2) and (3) are used. The admittance is the presentation of an equivalent parallel conductance (real part) G(ω) and capacitance C(ω) (imaginary part) [8].

$$Y^*(\omega) = \frac{I}{V} = i\omega C^*(\omega) \equiv G(\omega) + i\omega C(\omega) \quad (1)$$

C*(ω) and Y*(ω) is the complex capacitance and complex admittance

The complex impedance is the inverse of the complex admittance.

$$Z^*(\omega) = \frac{V}{I} = \frac{1}{Y^*} \quad (2)$$

While the dielectric constant is proportional to the complex capacitance.

$$C^*(\omega) = \frac{S}{e}\varepsilon^*(\omega) \quad (3)$$

And the complex conductivity is proportional to the admittance.

$$\sigma^*(\omega) = \frac{e}{S}Y^*(\omega) \quad (4)$$

*S* being the area of the electrode and *e* is the thickness of the film.

The use of different immittance allows the determination of different contributions (bulk of the film, electrode/film interface and/or grain boundaries) having very different dielectric response in amplitude and frequency.

## Experimental details

The $BiFe_{0.95}Mn_{0.05}O_3$ (BFMO) thin film (250 nm thickness) has been grown by pulsed laser deposition (PLD) using an excimer laser (KrF, wavelength 248nm) at a pulse repetition rate of 6 Hz, a target to substrate distance of 4.5 cm, 740 °C and 10 Pa pressure of oxygen. The fluence is 2 J/cm$^2$ and the total time deposition about 55 minutes. To prevent the Bi loss during the growth and to decrease the amount of leakage currents, a commercial target (from CERACO ceramic coating GmbH) with 10 % excess of bismuth and an iron substitution with 5 % of Mn has been used ($Bi_{1.1}Fe_{0.95}Mn_{0.05}O_3$) to prevent the leakage current [9]. Indium-Tin oxide (ITO) top electrodes (200 µm diameter) have been deposited by PLD through a mask at room temperature (1.6 Pa oxygen pressure). The bottom electrode $SrRuO_3$ (SRO) has been deposited by PLD at 710 °C temperature and 30 Pa of oxygen pressure. ITO and SRO ceramic targets were purchased from Neyco Vacuum & Materials. The ITO/BFMO/SRO heterostructure has been grown on (001) oriented $SrTiO_3$ (STO) substrate (common target to substrate distance of 4.5cm and laser fluency of 2 J/cm$^2$ ; STO substrate purchased at Crystal GmbH).

Dielectric measurements were performed using an impedance-meter (Solartron 1260A) with a broad range of frequency (1 Hz to $10^6$ Hz) and an alternative current (AC) excitation amplitude of 0.6 V (far below the coercive field, see figure 1). The direct current (DC) measurement was performed by a source-meter Keithley 2635B. Linkam stage was used to investigate the temperature influence (from 213 K to 353 K) on the dielectric response. Ferroelectric hysteresis loops were measured using an aicACCT TF analyser 3000. The X-ray diffraction pattern has been acquired by Siemens diffractometer D5000 ($K\alpha_1$ 1.54056 Å wavelength) in θ/2θ geometry.

## Results and discussions

Figure 1(a) presents X-ray diffraction pattern of the BFO thin film in θ/2θ geometry. Only intense $(00l)_{pc}$ reflexions are observed indicating an epitaxial growth. The out-of-plane lattice parameter calculated using the Bragg law for the BFO layer (3.99 Å) and SRO layer (3.95 Å) are higher than the pseudo-cubic lattice parameter of the bulk BFO and SRO. This suggests the presence of a compressive in plane strain due to the epitaxial mismatch with the STO substrate (3.905 Å). Rocking curves (ω scan) performed around the STO (001) and BFO $(00l)_{pc}$ reflexions indicate a good orientation of the BFO layer (full width at half maximum (FWHM) for (BFO) = 0.5° and FWHM(STO) = 0.07° not shown). A supplementary peak is detected in log scale in the Figure 1 (a) at 35° possibly due to a parasitic phase (most likely $Fe_2O_2$ [10]). Note that the intensity of the parasitic phase is about 1000 % lower compared to the BFO $(00l)_{pc}$ reflexion. The parasitic phase does not affect the electrical proprieties of our sample and a highly resistive and robust ferroelectric state is obtained as evidenced by the ferroelectric polarization hysteresis (Figure 1(b)).

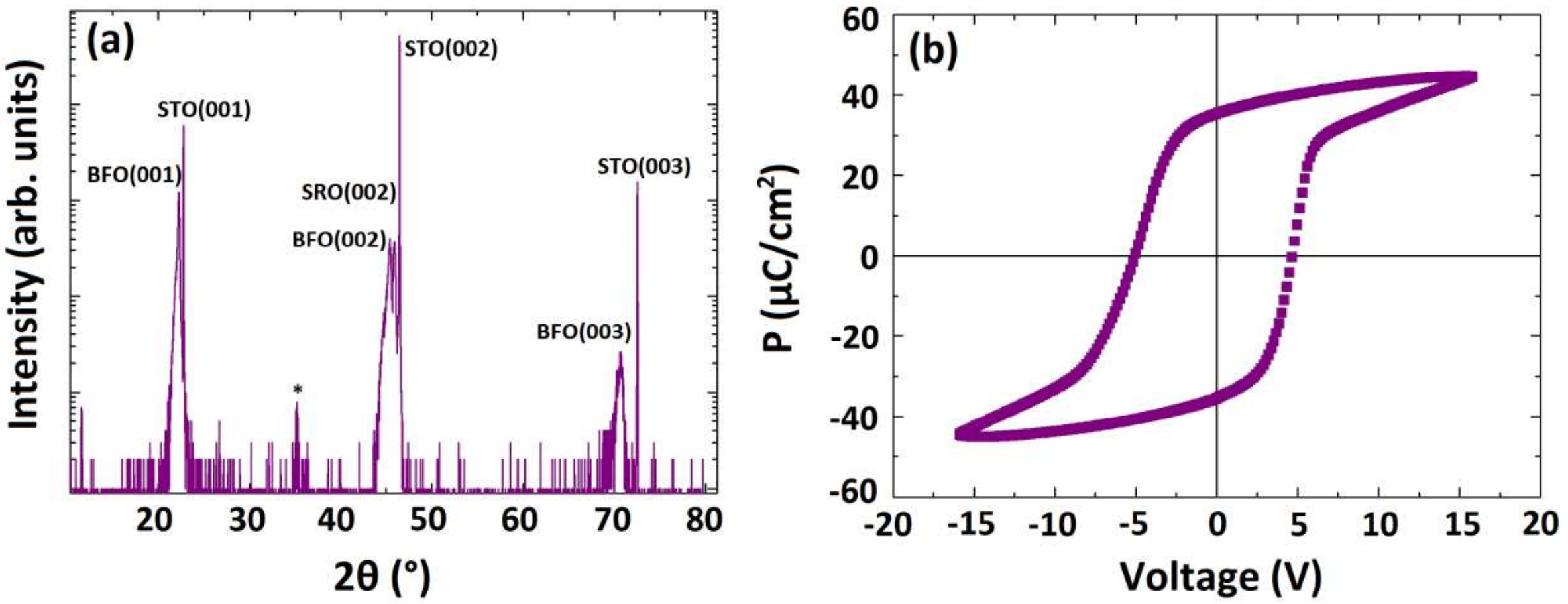

Figure 1. (a) X-ray diffraction pattern for the $BiFeO_3$ thin film, * Parasitic phase. (b) Ferroelectric hysteresis loop of the BFO thin film.

Figure 2 presents the complex impedance plot (*Z*'' *vs Z*') for a wide range of temperature (213 K to 353 K), with fitting result for each temperature. The impedance plot cannot be described by a perfect semi-circle which implies multiple contribution to the dielectric response. Dielectric response of the heterostructure is modelled using an equivalent circuit composed of two (RC) in series with an additional resistance r (figure 2 (b)). The first $(R_b,C_b)$ circuit corresponds to the bulk contribution while the second $(R_i,C_i)$ circuit describes the interface contribution. The resistance r corresponds to the electrode resistance and it is generally very small. From this equivalent circuit (figure 2(b)) an expression of the impedance can be given:

$$Z_{tot} = Z_b + Z_i + r \qquad (5)$$

$Z_b$ being the bulk impedance and $Z_i$ the interface impedance contribution.

$$Z_b = \frac{R_b}{1+jC_bR_b\omega} \text{ and } Z_i = \frac{R_i}{1+jC_iR_i\omega} \qquad (6)$$

$$Z_{tot} = \left(\frac{R_b}{1+(C_bR_b\omega)^2} + \frac{R_i}{1+(C_iR_i\omega)^2} + r\right) - j\left(\frac{C_bR_b^2\omega}{1+(C_bR_b\omega)^2} + \frac{C_iR_i^2\omega}{1+(C_iR_i\omega)^2}\right) \qquad (7)$$

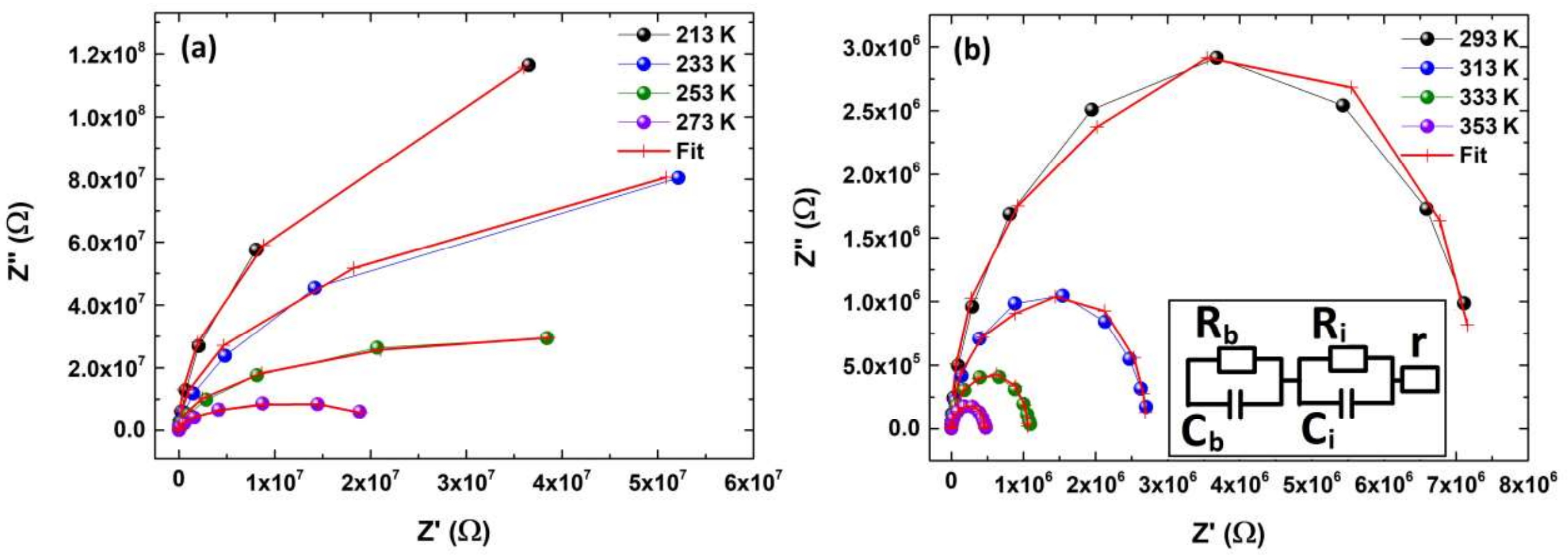

Figure 2. Complex impedance plan plot (Z'' vs Z') (a) 213 K to 273 K, (b) 293 K to 353 K. Insert in (b) shows the equivalent circuit chosen to model the impedance results. High and low temperatures are deliberately shown separately for clarity reason.

The fitting parameters are gathered in table 1. The value of ($R_b$,$C_b$) parameters are lower than ($R_i$,$C_i$) because contributions from interfaces tend to be higher than bulk in BFO systems. The fitted resistance $R_b$ and $R_i$ are decreasing with temperature (insulating behavior), while the evolution of the fitted capacitance $C_b$ and $C_i$ do not show any trend (far from Tc = 1100 K in bulk for BFO). The value of $C_b$ and $C_i$, are close (same magnitude) to those found in the literature [11-12].

| Temperature (K) | $C_b$ (F) | $R_b$ (Ω) | $C_i$ (F) | $R_i$ (Ω) | r (Ω) |
|---|---|---|---|---|---|
| 353 K | $2.18\ 10^{-10}$ | $1.59\ 10^{5}$ | $5.85\ 10^{-10}$ | $3.02\ 10^{5}$ | $5.55\ 10^{2}$ |
| 333 K | $2.46\ 10^{-10}$ | $3.81\ 10^{5}$ | $6.12\ 10^{10}$ | $6.77\ 10^{5}$ | $7.2\ 10^{2}$ |
| 313 K | $2.01\ 10^{-10}$ | $9.68\ 10^{5}$ | $6.37\ 10^{-10}$ | $1.73\ 10^{6}$ | $8.43\ 10^{2}$ |
| 293 K | $2.05\ 10^{-10}$ | $2.1\ 10^{6}$ | $4.62\ 10^{-10}$ | $5.17\ 10^{6}$ | $8.02\ 10^{2}$ |
| 273 K | $2.11\ 10^{-10}$ | $6.84\ 10^{6}$ | $4.6\ 10^{-10}$ | $1.43\ 10^{7}$ | $8.27\ 10^{2}$ |
| 253 K | $2.21\ 10^{-10}$ | $2.21\ 10^{7}$ | $4.2\ 10^{-10}$ | $4.81\ 10^{7}$ | $8.59\ 10^{2}$ |
| 233 K | $1.84\ 10^{-10}$ | $7.69\ 10^{7}$ | $3.63\ 10^{-10}$ | $2.34\ 10^{8}$ | $6.48\ 10^{2}$ |
| 213 K | $1.70\ 10^{-10}$ | $2.14\ 10^{8}$ | $4.28\ 10^{-10}$ | $8.31\ 10^{8}$ | $4.48\ 10^{2}$ |

Table 1. Fitting parameters for several temperature (213-353 K), obtained from the equivalent circuit [($R_bC_b$)($R_iC_i$)(r)].

To identify the bulk and interface contribution, we present in figure 3 the frequency dependence of Z' and Z" for different temperatures (213 K to 353 K). We observe in this figure an impedance relaxation with frequency, strongly dependent on temperature. In figure 3(a) on heating the real part Z' becomes independent of frequency. The range of frequency where Z' is constant becoming wider on further heating the temperature. Simultaneously a relaxation peak in the imaginary part Z" emerges (figure 3(b)). On heating this peak is shifted toward high frequency from about 10 Hz to 1 kHz. We attribute this relaxation process to the interface contribution (as confirmed below from the comparison with the DC conduction mechanism). At low temperature the interface contribution being minor compared to the bulk contribution. The evolution of the relaxation peak position with temperature is a signature of a thermally activated relaxation process. The time relaxation ($\tau = \frac{1}{2\pi f_p}$, $f_p$ is the relaxation peak maximum frequency) of this process follows an Arrhenius law $\tau = \tau_0 exp\left(\frac{E_a}{k_B T}\right)$, as we can see in the insert of figure 3(b). $E_a$ represents the energy activation of the process. The time relaxation can also be expressed as $\tau$=$R_iC_i$, where $R_i$ and $C_i$ being the interface resistance and capacitance of the (R,C) model in figure 2(b). An activation energy $E_a$ value of 0.39 eV is extracted from the linear fit (see insert Fig. 3(b)).

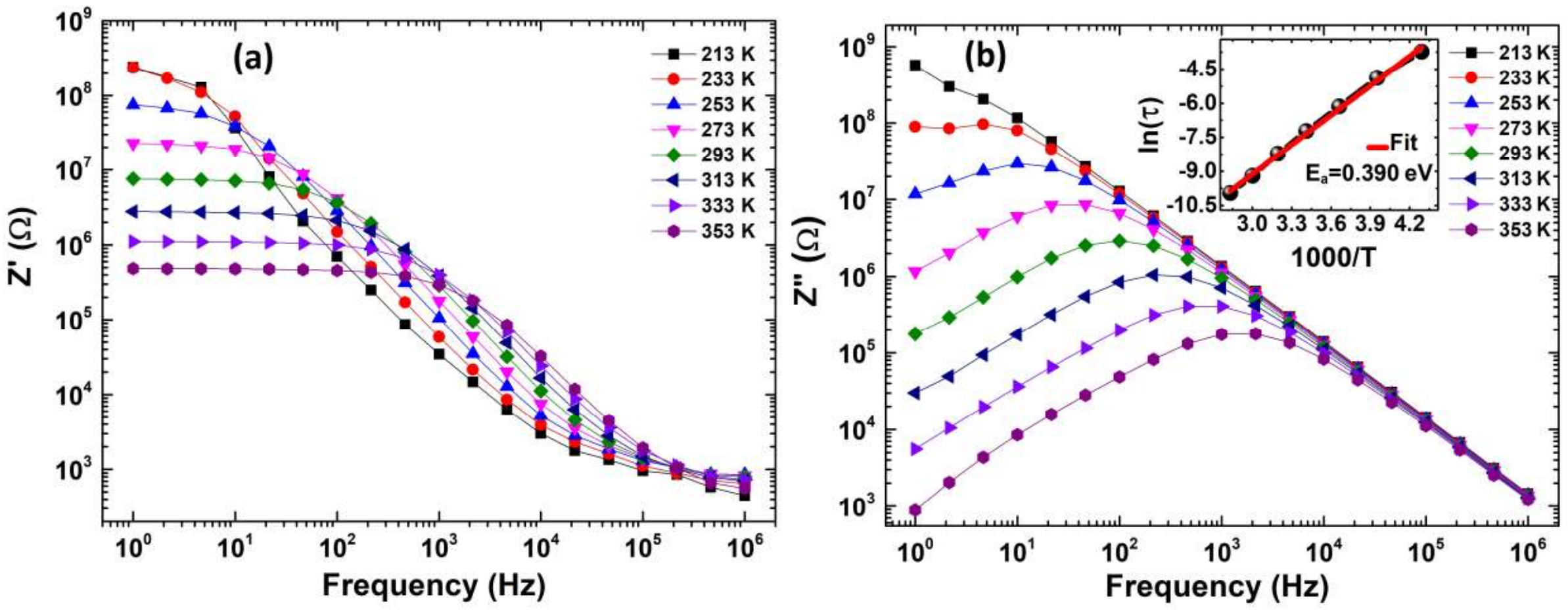

Figure 3. Frequency dependence of (a) the real part Z' and (b) the imaginary part Z'' at different temperatures (213 K to 353 K). Insert in (b) shows the Arrhenius representation of the relaxation time.

Figure 4 presents the real and imaginary part (ε' and ε'' respectively) of the dielectric function as a function of frequency for several temperatures. Recall that ε' and ε'' have been calculated from equations (1), (2) and (3). Figure 4 shows a difference in the behaviour of ε' and ε'' between low and high temperature. Indeed ε' is constant over a large range of frequency at low temperature (figure 4 (a)) while it presents larger values for low frequencies on increasing temperatures. This thermally activated contribution to ε' become larger and extends to high frequency on heating up the sample. This contribution is quite common in oxide heterostructures and is due to space charge layer at the interface between the BFO film and the electrodes (ITO/BFO and BFO/SRO), also known as Maxwell-Wagner effect. At low temperature (figure 4(a) and (b)) the Maxwell-Wagner effect is not observed and we suspect that the interface effect at low temperature is somehow frozen. Only the bulk contribution to the permittivity seems to be present. The intrinsic static dielectric constant show values below 100 as published by other groups and similar values are measured in figure 4 (a) (but here it is observed at high frequency). Values up to 700 are observed on figure 4 at high temperature and low frequencies confirming an extrinsic origin. The Maxwell-Wagner effect also has an impact on the dielectric loss ε''. The low frequency dielectric loss increases on heating as shown in figure 4 (b,d,f). Such space-charge induced loss is probably superposed to the leakage current (low frequency). We also note an increase of the dielectric loss ε'' at high frequency and we suppose a ferroelectric dipole origin for this relaxation [8][13]. This point needs further investigation. From this dielectric investigation, a contribution of the space charge interface is therefore evidenced in agreement with the relaxation observed from the impedance spectra and the interface limited conduction mechanism.

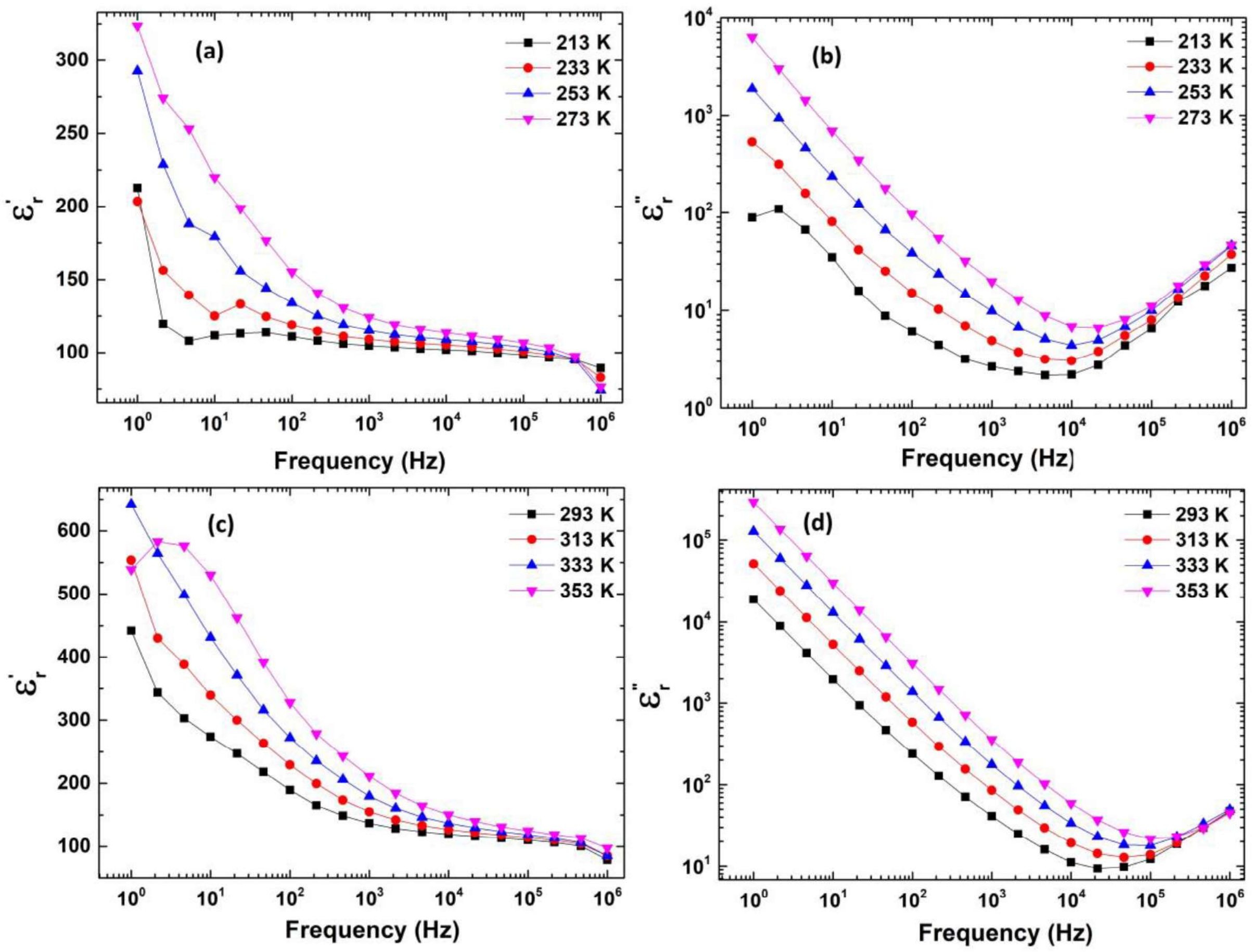

Figure 4. The evolution of the permittivity (real and imaginary part) with frequency for two temperature ranges.

To better understand this interfacial relaxation process and the possibility of other contributions, a study of the AC conduction is performed. We recall that combining the relations (1), (3) and (4) the real part of the complex conductivity is proportional to the imaginary part of the dielectric constant $\varepsilon^*$ which corresponds to the dielectric loss ($\varepsilon'' \approx \sigma'$ ). We would like to stress out again that the use of different representations and immittance allows disentangling the different dielectric contribution within the heterostructure. Figure 5(a) shows the temperature and frequency evolution of the real part of the complex conductivity, three distinct zones can be distinguished. The first one is frequency independent and begins at low frequency corresponding to the DC contribution. The second zone is frequency dependent and appears at high frequency and corresponds to the AC contribution. Such AC contribution is characterized by an increase in the σ' conductivity on increasing frequency. The two zones evolve with temperature. At low temperature the plateau is almost absent in the frequency window. The DC plateau becomes, however, larger on increasing temperature (figure 5 (a)). The high frequency limit of the DC plateau reaches for instance $10^4$ Hz at high temperature (353 K).

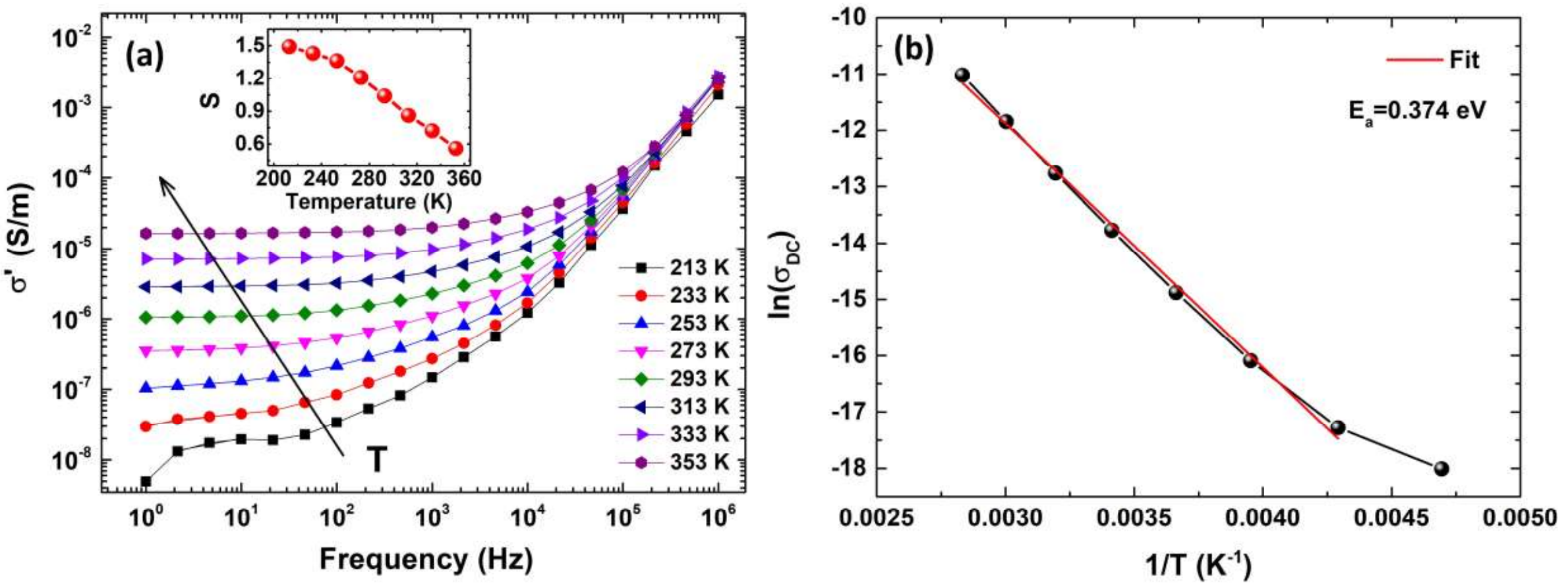

Figure 5. (a) Frequency dependence of the real part of the complex conductivity for several temperatures (213 K to 353 K) in logarithmic presentation. Insert: Exponent S of Jonscher's law versus temperature. (b) Temperature dependence of the DC conductivity extracted from extrapolation to zero frequency. The linear fit of the Arrhenius plot gives a thermally activated energy of $E_a = 0.374$ eV.

The extrapolation to zero frequency of the plateau conductivity on Figure 5 (a) allows the estimation and discussion of $\sigma_{DC}$. A simple look at the (R,C) model (figure 2(b)) shows that the DC conductivity is the conduction through the resistive part (Ri+Rb+r). The Arrhenius plot on Figure 5(b) clearly suggests a thermally activated process for the DC conduction with an energy $E_a = 0.374$ eV. A comparison will be made below with the DC conductivity investigation. Before proceeding to the DC properties and analysis, we would like to point an interesting behaviour observed on Figure 5 (a) for the high frequencies. We observe an overlap of the curves corresponding to different temperatures into a universal power law. The AC conduction in this power law regime is due to the charge carrier hopping [13]. The Jonscher law of such charge hopping is often used in oxides and is given by the following relation $\sigma' = \sigma_{DC} + A\omega^S$ [13-15]. We fitted the conductivity part depending on the frequency with the power law (Jonscher law) $A\omega^S$. The exponent s is presented in the insert in Figure 5 (a). It decreases with temperature from 1.49 (213 K) to 0.56 (353 K). The value of the exponent s is related to the hopping type : translational ($s<1$) or localized/reorientational ($s>1$) [16]. The decrease of the exponent s with temperature can be explained by two mechanisms: polaron tunnelling or correlated barrier hopping (CBH) [17][18]. Generally, tunnelling process is favored at low temperature, which suggests a CBH mechanism for the behaviour of the AC conductivity in Figure 5. Additional investigation using a wider frequency range and temperature are needed to conclude on the exact hopping mechanism at high frequencies.

Combining the above representations and analysis (conductivity, impedance and permittivity) we identify therefore two contributions to the dielectric response: the contribution of the interfaces (space charge layers) at low frequencies and high temperatures and the bulk of the film at higher frequencies and lower temperatures (suggested by the power law or Jonscher law behaviour of the conductivity). The origin of the space charge layer (Maxwell-Wagner effect) might affect the DC transport mechanism as suggested by the DC extrapolation of $\sigma_{DC}$ (figure 5) and we thus switch to the I(V) analysis.

Figure 6 shows the evolution of the absolute value of the current density with voltage for multiple temperatures (213 K-353 K). The asymmetric current density for the positive and negative voltage (figure 6) is a signature of an interface limited conduction in the thin film. Different models are observed in dielectric thin films for interface limited transport. The Fowler-Nordheim mechanism is excluded considering the observed temperature dependence. The most observed and investigated is the Schottky mechanism [14-15] and a derived Schottky-Simmons emission model gives excellent agreement with our thin film transport proprieties.

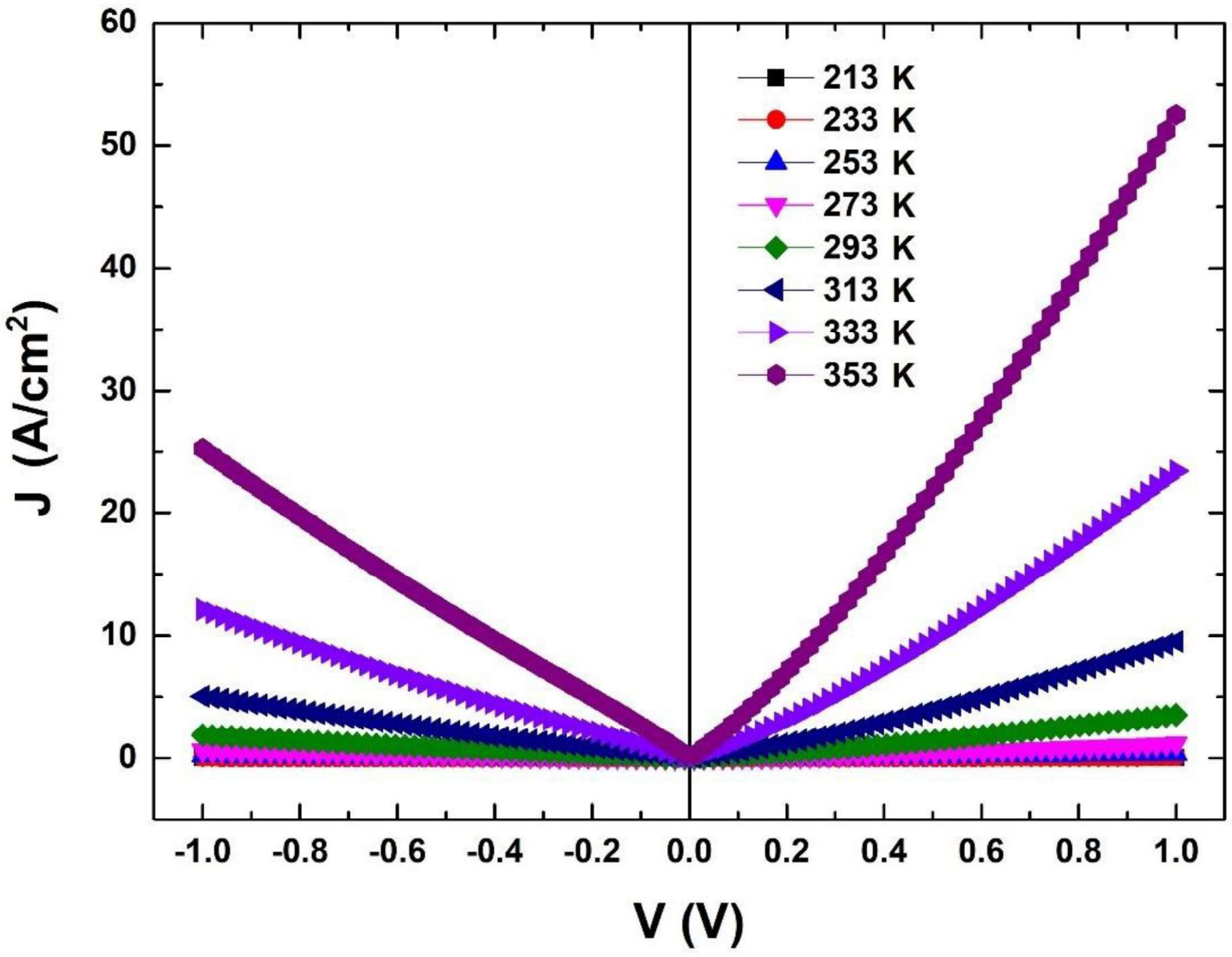

Figure 6. Current density evolution with voltage for multiple temperature (213 K-353 K).

From the Schottky-Simmons model, the current density depends exponentially in temperature and electric field [19].

$$J = 2q\mu E\left(\frac{2\pi m_e k_B T}{h^2}\right)^{\frac{3}{2}} exp\left(-\frac{q}{k_B T}\left(\phi_B^0 - \sqrt{\frac{qE}{4\pi\varepsilon_0\varepsilon_{HF}}}\right)\right)$$

The linear form of this equation is used to fit the experimental data.

$$ln\left(\frac{J}{T^{\frac{3}{2}}}\right) = ln\left(2q\mu E\left(\frac{2\pi m_e k_B}{h^2}\right)^{\frac{3}{2}}\right) - \frac{q}{k_B}\left(\phi_B^0 - \sqrt{\frac{qE}{4\pi\varepsilon_0\varepsilon_{HF}}}\right)\frac{1}{T}$$

This equation is identified to a linear equation y=ax+b.

With $a = -\frac{q}{k_B}\phi_{app}$ and $\phi_{app} = \phi_B^0 - \left(\sqrt{\frac{q}{4\pi\varepsilon_0\varepsilon_{HF}}}\right)\sqrt{E}$

The Schottky potential height is extracted from the fit of the plot in the figure 7. The calculated potential barrier ($\phi_B^0$ = 0.360 eV) is very close to the activation energy found above (see figure 5). The obtained low value of the dielectric constant $\varepsilon_{HF}$ = 24 is in agreement with the Schottky Simmons model which assumes a high frequency dielectric constant for the model (the time travel of the charge through the barrier being short, a high frequency dielectric constant is taken into account ; the optical dielectric constant is close to 5.4 for BFO).

From the equation above, b= $ln\left(2q\mu E\left(\frac{2\pi m_e k_B}{h^2}\right)^{\frac{3}{2}}\right)$ but the mobility μ and effective mass $m_e$ are unfortunately unknown for this BFO thin film. Simple estimates and discussion can however be made. Indeed, from the literature the effective mass for several perovskite is about 4 to 5 times the free electron mass [20-22]. To extract the mobility value, we assume therefore that the effective mass is $m_e$=4$m_0$. The calculated electrical mobility is then μ = 0.02 $cm^2.V^{-1}.s^{-1}$ .This value is lower than the drift mobility value typically found in literature (0.1-0.5 $cm^2.V^{-1}.s^{-1}$) for perovskites but can be explained by the presence of disorder (oxygen and bismuth vacancies and manganese doping on the B-site )[22,23-24].

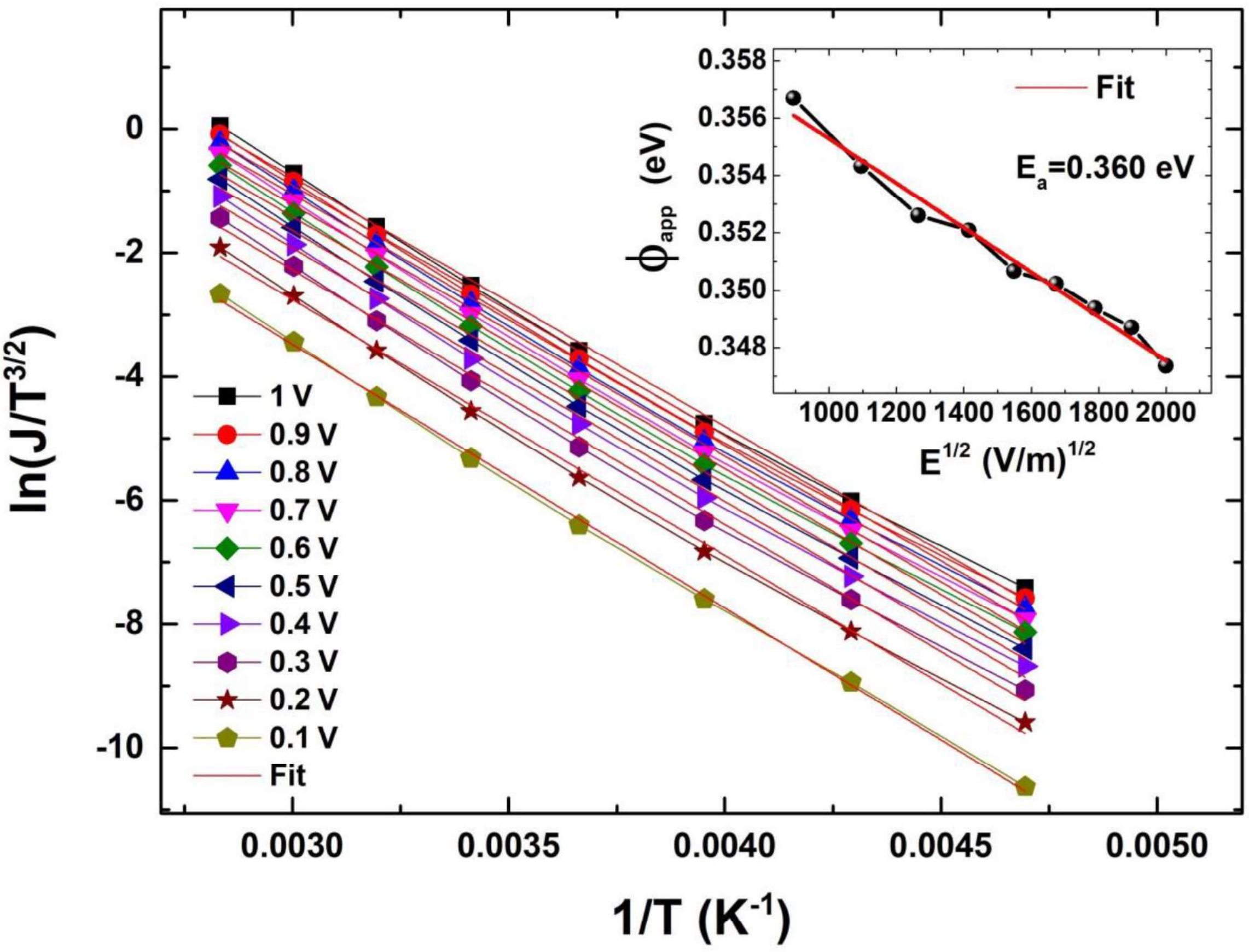

Figure 7. ln($J/T^{3/2}$) vs 1/T plot for the Schottky-Simmons model at different temperatures. Insert: Potential versus Field to deduce the Schottky barrier.

## Conclusion

A detailed investigation of the dielectric response of a BFO thin film grown on (001) oriented $SrTiO_3$ substrate is provided with a parallel plate capacitance geometry (top and bottom electrode respectively ITO and SRO). Measurements at different frequencies and temperatures combined with an equivalent circuit model allow detecting two contributions to the dielectric response from the interfaces and the bulk of the thin film. The use of different immittances also helped into disentangling the bulk versus interface contributions both having specific amplitudes, temperature and frequency ranges. The thermally activated (Ea = 0.37 eV) interface relaxation of Maxwell-Wagner type seems to be closely related to the Schottky space charge as confirmed by the very close activation energy value obtained from the Schottky-Simmons Model (Ea = 0.36 eV). Based on the Schottky-Simmons model of the DC transport we also tentatively provide estimates of the drift mobility and the estimated low value is interpreted as due to disorder (vacancies, dislocations, grain boundaries). Finally, further investigations are clearly needed to better understand the bulk contribution and confirm the polaronic hopping mechanism deduced from the high frequency AC conductivity behaviour. Polaronic mechanism has already been detected in BFO and Bismuth based complex oxides and seems to be at the heart of the exotic photo-induced properties in BFO and justify therefore additional research [25-27].

**Acknowledgement**

The simulation is the results of using the program ZfitGui.m developed by Dr Jean-Luc Dellis. The programme is free and accessible from the Matlab store. Financial support by H2020-RISE-ENGIMA-778072 and the region of Picardy (project ZOOM) are gratefully acknowledged.

Tables (if any)

| Temperature (K) | $C_b$ (F) | $R_b$ (Ω) | $C_i$ (F) | $R_i$ (Ω) | r (Ω) |
|---|---|---|---|---|---|
| 353 K | $2.18\ 10^{-10}$ | $1.59\ 10^{5}$ | $5.85\ 10^{-10}$ | $3.02\ 10^{5}$ | $5.55\ 10^{2}$ |
| 333 K | $2.46\ 10^{-10}$ | $3.81\ 10^{5}$ | $6.12\ 10^{10}$ | $6.77\ 10^{5}$ | $7.2\ 10^{2}$ |
| 313 K | $2.01\ 10^{-10}$ | $9.68\ 10^{5}$ | $6.37\ 10^{-10}$ | $1.73\ 10^{6}$ | $8.43\ 10^{2}$ |
| 293 K | $2.05\ 10^{-10}$ | $2.1\ 10^{6}$ | $4.62\ 10^{-10}$ | $5.17\ 10^{6}$ | $8.02\ 10^{2}$ |
| 273 K | $2.11\ 10^{-10}$ | $6.84\ 10^{6}$ | $4.6\ 10^{-10}$ | $1.43\ 10^{7}$ | $8.27\ 10^{2}$ |
| 253 K | $2.21\ 10^{-10}$ | $2.21\ 10^{7}$ | $4.2\ 10^{-10}$ | $4.81\ 10^{7}$ | $8.59\ 10^{2}$ |
| 233 K | $1.84\ 10^{-10}$ | $7.69\ 10^{7}$ | $3.63\ 10^{-10}$ | $2.34\ 10^{8}$ | $6.48\ 10^{2}$ |
| 213 K | $1.70\ 10^{-10}$ | $2.14\ 10^{8}$ | $4.28\ 10^{-10}$ | $8.31\ 10^{8}$ | $4.48\ 10^{2}$ |

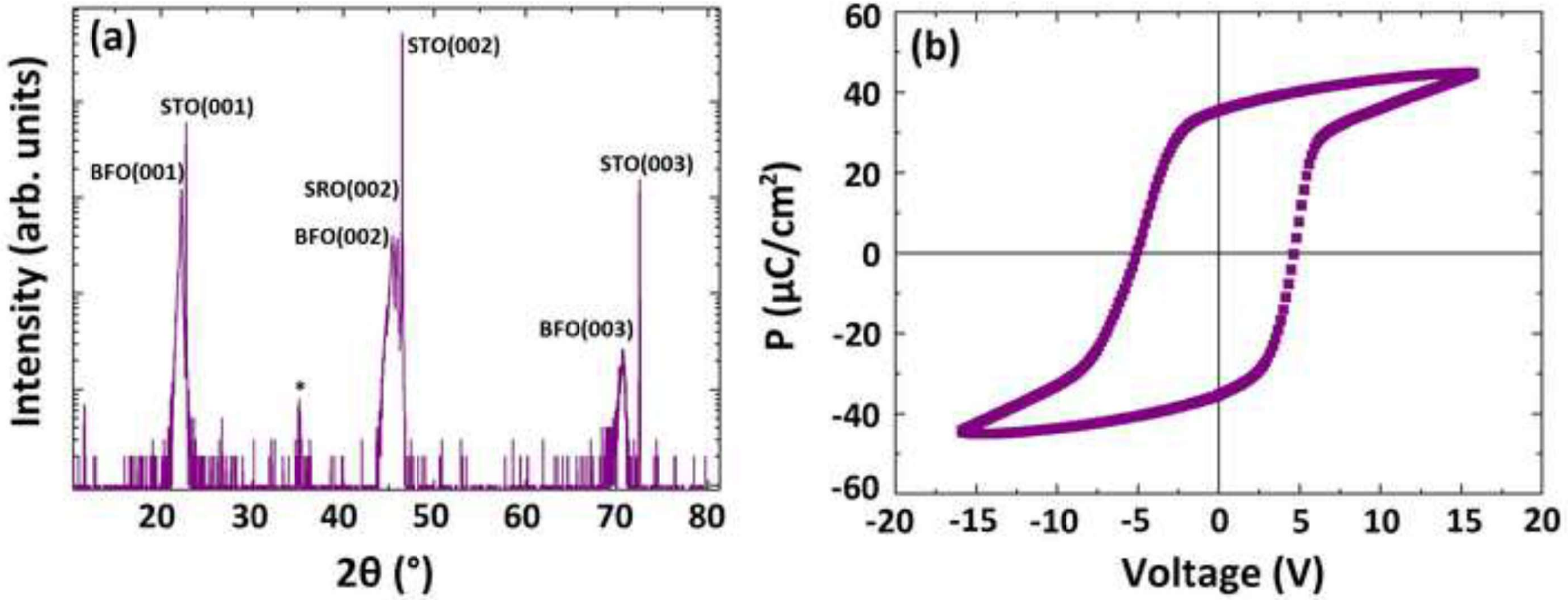

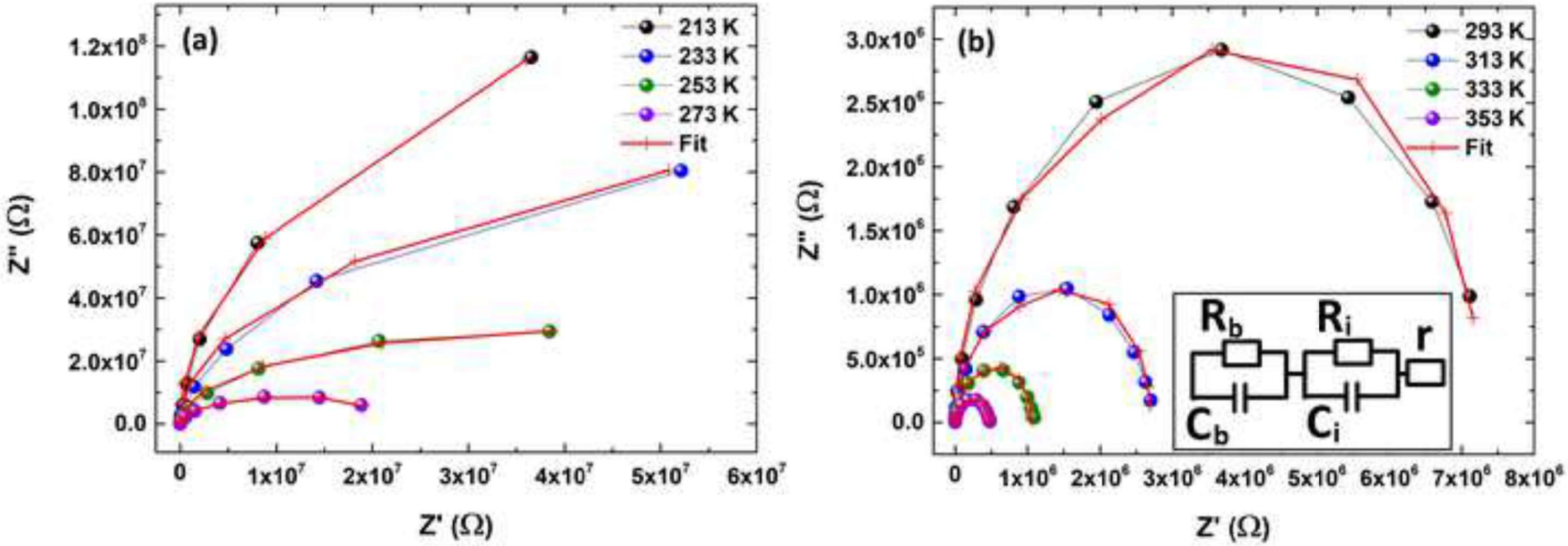

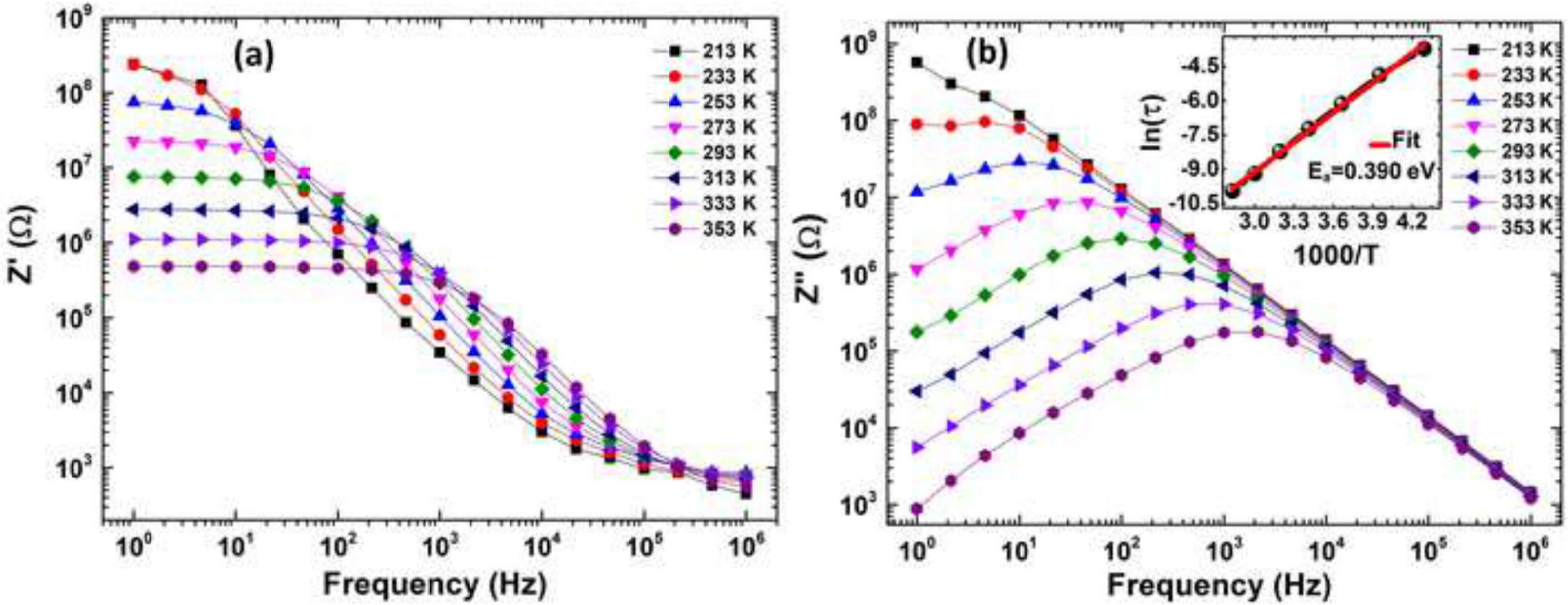

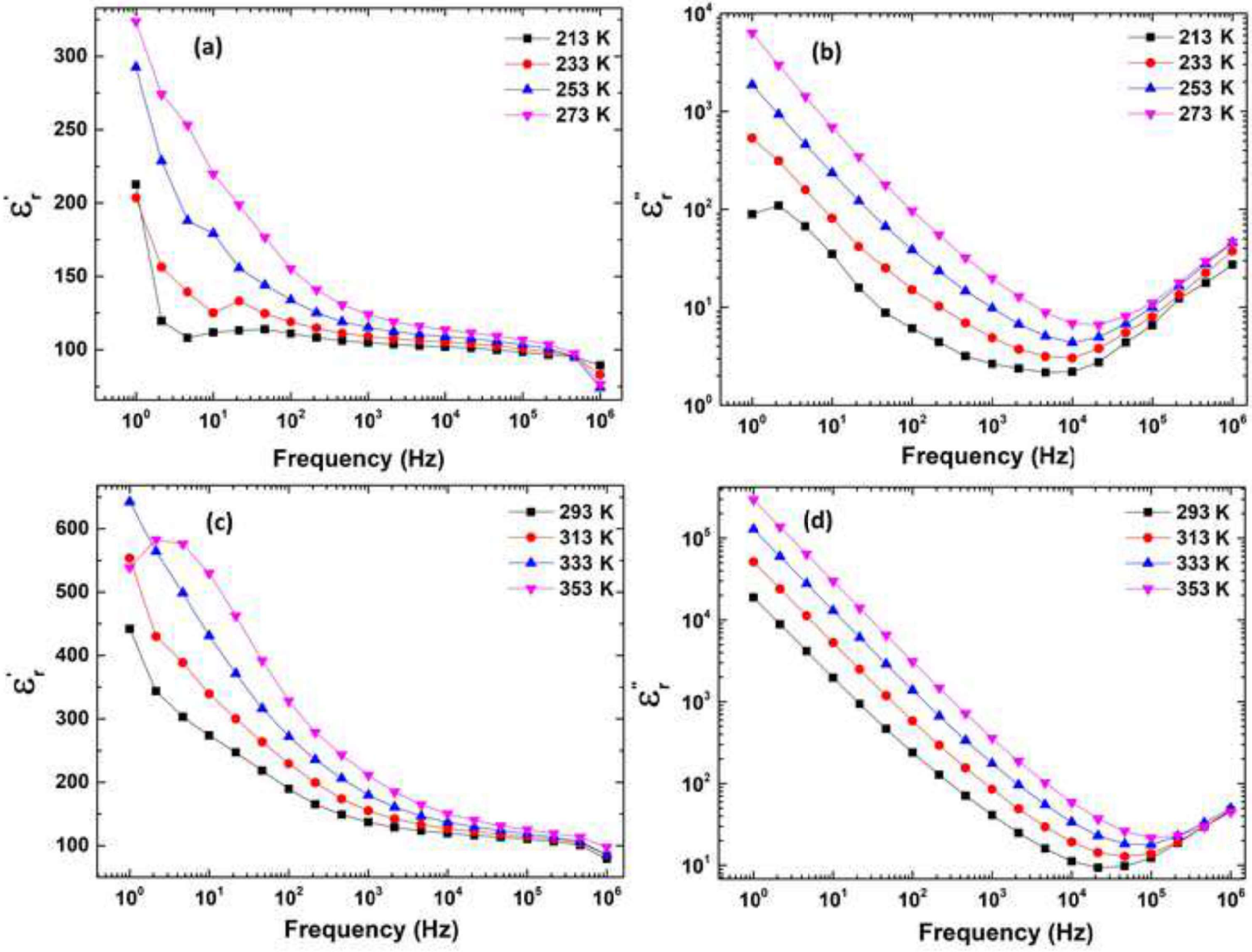

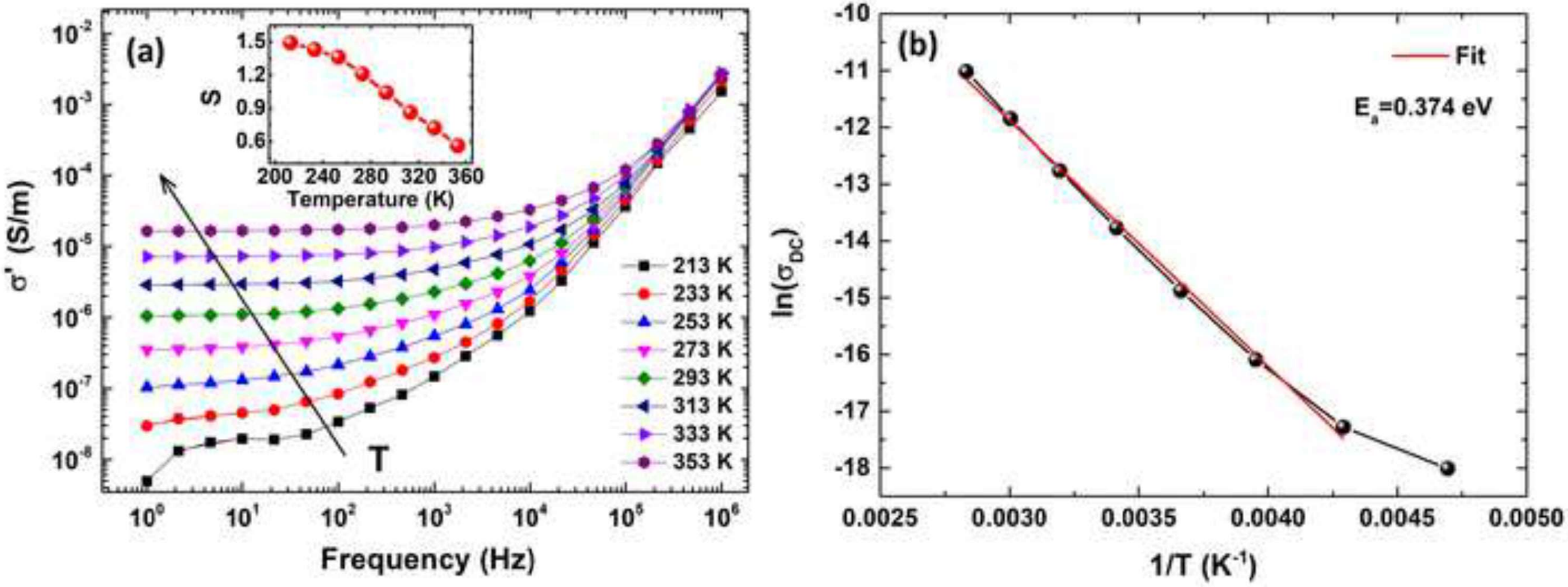

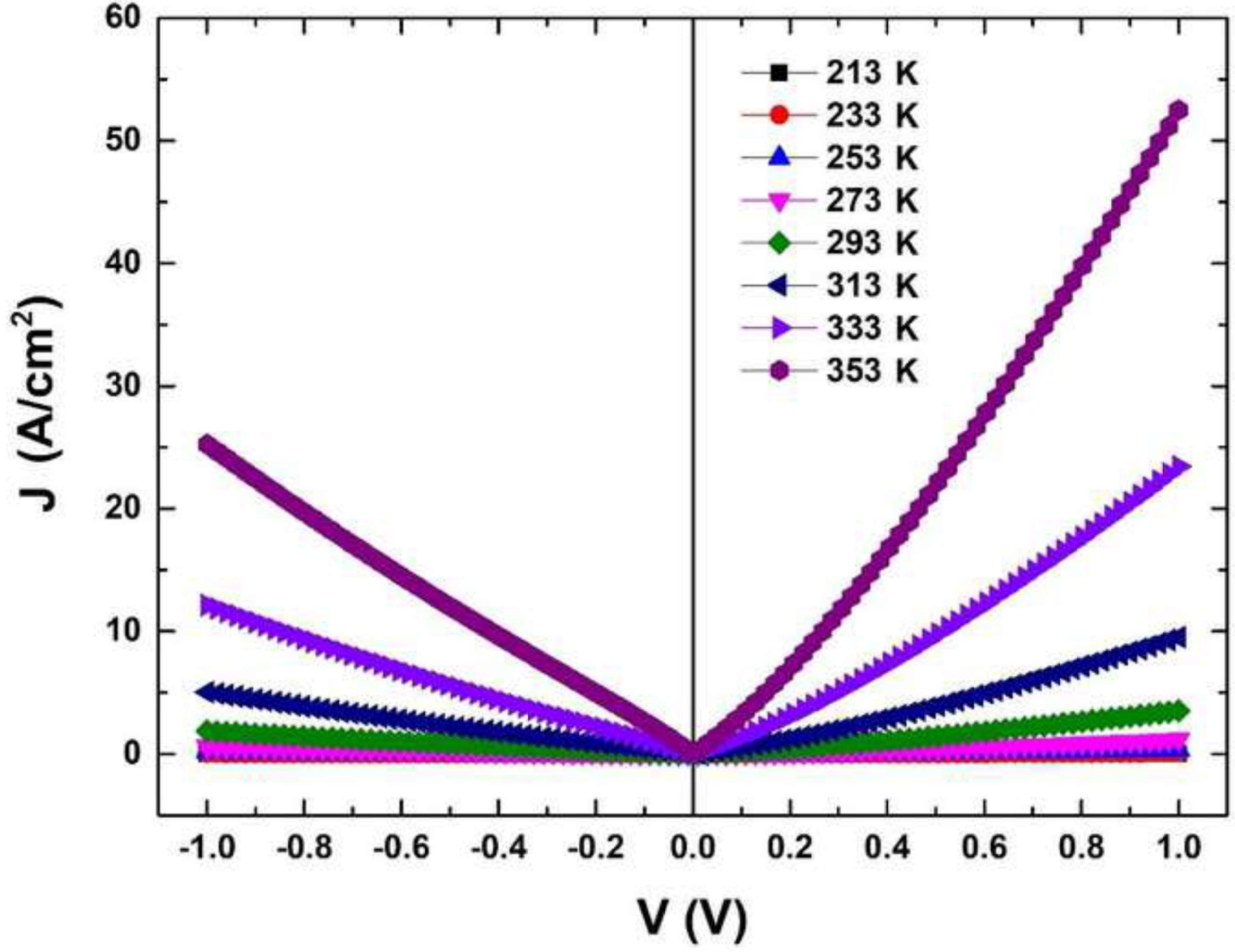

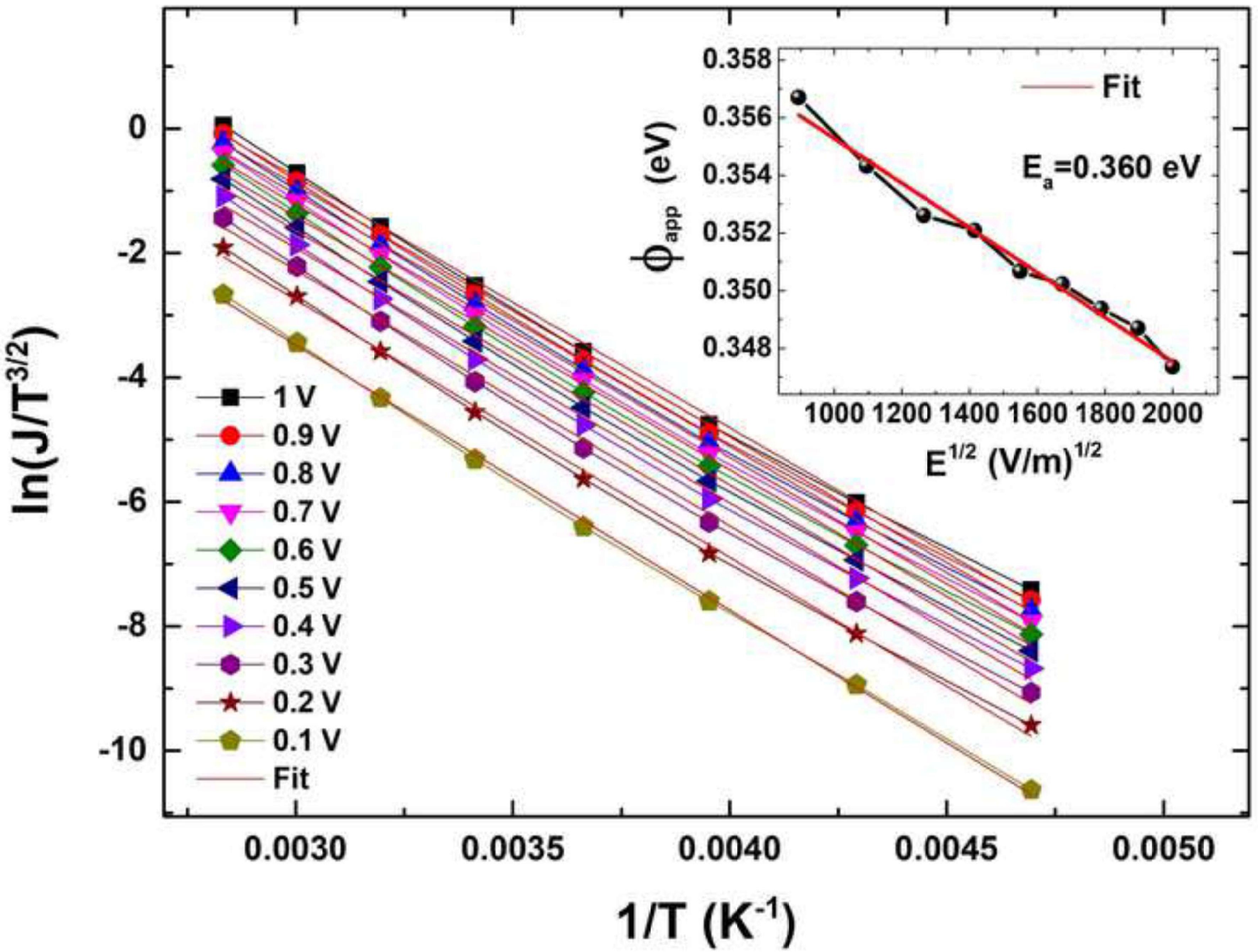